\documentclass[lettersize,journal]{IEEEtran}

\usepackage{amsmath,amsfonts}
\usepackage{algorithm}
\usepackage{array}
\usepackage{textcomp}
\usepackage{stfloats}
\usepackage{url}
\usepackage{verbatim}
\usepackage{graphicx}
\usepackage{cite}

\usepackage[utf8]{inputenc}
\usepackage{ifthen}
\usepackage{hyperref}
\usepackage[inline]{enumitem} 
\usepackage[dvipsnames,table]{xcolor} 
\usepackage{soul}
\usepackage[normalem]{ulem} 
\usepackage{subcaption}
\usepackage{epsf,picinpar}
\usepackage{varioref}
\usepackage{algorithm}
\usepackage[noend]{algpseudocode} 
\usepackage{fdsymbol}
\usepackage{tcolorbox}
\usepackage{forest} 
\usepackage{enumerate}
\usepackage[framemethod=TikZ]{mdframed}

\usepackage{pifont}
\usepackage[pscoord]{eso-pic}

\usepackage{booktabs}
\usepackage{multirow}
\usepackage{float} 
\usepackage{harveyballs}
\usepackage{makecell}
\usepackage{wrapfig}
\usepackage{svg}
\usepackage{orcidlink}
\usepackage{listings}
\usepackage{tikz}
\usepackage{calc} 
\usepackage{framed}

\newcommand{\secref}[1]{Section~\ref{#1}}

\newcommand{\figref}[1]{Fig.~\ref{#1}}
\newcommand{\algoref}[1]{Algorithm~\ref{#1}}
\newcommand{\tabref}[1]{Table~\ref{#1}}

\newcommand{\todo}[1]{
    \ifthenelse{\boolean{showannotations}}%
    {\ifthenelse{\equal{#1}{}}{\textcolor{red}{TODO}}{\textcolor{red}{TODO:~{#1}}}}%
    {}%
}

\newcommand{\assignedto}[1]{%
    \ifthenelse{\boolean{showannotations}}%
    {\textbf{\noindent\ding{46}\textcolor{white}{~\colorbox{\assignementcolor}{Assigned to:}}~\textcolor{\assignementcolor}{#1}\\}%
    }
    {}
}

\newcommand{\rem}[1]{%
    \ifthenelse{\boolean{showannotations}}%
    {\textcolor{\oldtextcolor}{\st{#1}}}%
    {}%
}

\newcommand\add[1]{%
    \ifthenelse{\boolean{showannotations}}%
    {\textcolor{\newtextcolor}{{#1}}}%
    {#1}%
}

\newcommand\rep[2]{%
    \ifthenelse{\boolean{showannotations}}%
    {\rem{#1}~\add{#2}}%
    {#2}%
}

\newcommand{\darkmodecontrol}{%
    \ifthenelse{\boolean{darkmode}}%
    {\pagecolor{black!20}
    \color[rgb]{0,0,0} 
    }%
    {}%
}

\newcolumntype{L}[1]{>{\raggedright\let\newline\\\arraybackslash\hspace{0pt}}m{#1}}
\newcolumntype{C}[1]{>{\centering\let\newline\\\arraybackslash\hspace{0pt}}m{#1}}
\newcolumntype{R}[1]{>{\raggedleft\let\newline\\\arraybackslash\hspace{0pt}}m{#1}}

\floatstyle{ruled}
\newfloat{algo}{tbp}{loa}
\floatname{algo}{Algorithm}

\algnewcommand\algorithmicswitch{\textbf{switch}}
\algnewcommand\algorithmiccase{\textbf{case}}
\algnewcommand\algorithmicassert{\texttt{assert}}
\algnewcommand\Assert[1]{\State \algorithmicassert(#1)}%
\algdef{SE}[SWITCH]{Switch}{EndSwitch}[1]{\algorithmicswitch\ #1\ }{\algorithmicend\ \algorithmicswitch}%
\algdef{SE}[CASE]{Case}{EndCase}[1]{\algorithmiccase\ #1}{\algorithmicend\ \algorithmiccase}%
\algtext*{EndSwitch}%
\algtext*{EndCase}%
\algrenewcommand\algorithmicindent{1em}

\newtcolorbox{answerbox}{
  colback=gray!10,     
  colframe=black,      
  boxrule=0.3pt,       
  arc=3pt,             
  width=\columnwidth,
  boxsep=3pt,          
  left=3pt,            
  right=3pt,           
  top=3pt,             
  bottom=3pt           
}

\newboolean{showannotations}
\setboolean{showannotations}{true} 

\newboolean{shownames}
\setboolean{shownames}{true}

\newboolean{darkmode}
\setboolean{darkmode}{false} 

\newcommand{\newtextcolor}{blue}
\newcommand{\oldtextcolor}{red}
\newcommand{\assignementcolor}{orange}
\definecolor{highlightcolor}{rgb}{.99, 1, .0}
\sethlcolor{highlightcolor}
\definecolor{orcidlogocol}{HTML}{A6CE39}

\mdfsetup{%
   backgroundcolor=gray!3,
   roundcorner=3pt%
}

\newlength\WIDTHOFBAR
\lstdefinestyle{sqlstyle}{
    language=SQL,
    basicstyle=\ttfamily\small,
    keywordstyle=\color{blue}\bfseries,
    stringstyle=\color{orange},
    commentstyle=\color{gray}\itshape,
    morekeywords={MATCH,RETURN,UNWIND,CALL,YIELD,DETACH,WITH,OPTIONAL,to}, 
    showstringspaces=false,
    frame=single,
    columns=flexible,
    breaklines=true,
    xleftmargin=4mm,
    xrightmargin=4mm
}

\lstdefinelanguage{Kotlin}{
  morekeywords={package,import,class,interface,fun,object,val,var,typealias,
    is,in,Int,Double,Float,Boolean,String,Unit,Nothing,List,enum},
  sensitive=true,
  morecomment=[l]{//},
  morecomment=[s]{/*}{*/},
  morestring=[b]",
}

\definecolor{codegray}{gray}{0.35}
\definecolor{codegreen}{RGB}{40,120,40}
\definecolor{codeblue}{RGB}{40,70,150}

\lstdefinestyle{pythonstyle}{
    language=Python,
    basicstyle=\ttfamily\scriptsize,
    keywordstyle=\color{codeblue},
    commentstyle=\color{codegray},
    stringstyle=\color{codegreen},
    columns=fullflexible,
    keepspaces=true,
    showstringspaces=false,
    breaklines=false,
    aboveskip=0pt,
    belowskip=0pt
}
  
\begin{document}

\title{Type-IV Code Clone Detection via Layer-Wise Non-Contrastive Representation Learning}
\author{
\IEEEauthorblockN{
Luciano Marchezan,
Kevin Delcourt,
Eugene Syriani, and
Houari Sahraoui
}
\IEEEauthorblockA{\\
Université de Montréal. Montreal, QC, Canada\\
Email: \{luciano.augusto.marchezan.de.paula, kevin.delcourt\}@umontreal.ca}\\
\{syriani, sahraouh\}@iro.umontreal.ca
}

\markboth{Under review, September~2026}%
{Marchezan \MakeLowercase{\textit{et al.}}: Type-IV Code Clone Detection via Layer-Wise Non-Contrastive Representation Learning}

\maketitle
\begin{abstract}
Software clones are fragments of code that are similar or functionally equivalent to each other. They pose significant challenges for maintenance, refactoring, and bug detection. Detecting Type-IV clones, which are semantically equivalent but may differ syntactically, is particularly difficult for traditional token- or syntax-based methods. Recent machine learning approaches rely on contrastive learning, which requires careful negative sampling and can introduce bias. In this paper, we propose LWVIC4Code, a non-contrastive representation learning approach specifically designed for Type-IV clone detection. Building on the Variance-Invariance-Covariance Regularization (VICReg) framework and prior layer-wise VICReg training, LWVIC4Code introduces cross-layer consistency regularization and depth-dependent layer weighting to progressively refine semantic information across transformer layers, producing robust and discriminative code representations. We conduct an empirical study comparing LWVIC4Code against a contrastive learning baseline and zero-shot large language models on Python (Kamino) and multi-language (GPTCloneBench) datasets. Results show that LWVIC4Code achieves competitive or superior performance without negative samples, benefits from layer-wise supervision, and generalizes effectively from Python to other languages, particularly Java and C\#. These results demonstrate that non-contrastive, layer-wise representation learning is a promising direction for robust semantic code clone detection.
\end{abstract}

\section{Introduction}\label{sec:introduction}
Software clones, or similar fragments of code within or across software systems, are a common phenomenon in software development~\cite{Ain2019}. Detecting clones is critical for maintenance tasks such as code reuse, refactoring, bug detection, and quality assurance~\cite{Tairas2012}. For instance, over 50\% of files in two major open-source autonomous driving projects contain code clones, a substantial fraction of which are associated with previously reported bugs~\cite{mo2023comprehensive}.


Code clones are typically categorized into four types~\cite{Carter1993,Chanchal2009}: Types~I–III represent exact or near-identical fragments with syntactic similarity, while Type-IV clones are \emph{semantic clones}, i.e., functionally equivalent fragments that may differ entirely in syntax or structure. Detecting Type-IV clones requires reasoning about program semantics rather than surface-level syntactic patterns, which limits the effectiveness of traditional text- or token-based methods, especially when implementations differ in control-flow strategies, library usage, or algorithmic approach~\cite{Wang2023,Gabel2008,Dou2026}.

In this context, the growing adoption of large language models (LLMs) for code generation further amplifies the importance of clone detection, particularly for Type-IV clones. Recent studies indicate that AI-assisted programming is rapidly becoming standard practice in modern development workflows~\cite{Murali2024}. While LLMs improve productivity, they also introduce new challenges as the generated code may replicate patterns, logic, or entire solutions seen during training, sometimes closely resembling existing open-source implementations~\cite{Xu2025}. Such duplication is often not syntactically identical but semantically equivalent, making it difficult to detect using traditional clone detection techniques. This raises concerns related to code maintainability, redundancy, and licensing compliance when generated code inadvertently mirrors copyrighted sources~\cite{Wang2025}. Consequently, robust Type-IV clone detection is increasingly essential for identifying such semantic duplication across diverse implementations, programming styles, and programming languages.

Early research on semantic clone detection explored behavioral equivalence via program transformations~\cite{Gabel2008}, whereas more recent studies leverage machine learning (ML) to learn semantic representations of code~\cite{Bhaskar2024,Sheneamer2016}. Pretrained code embedding models such as CodeBERT~\cite{Wang2021} and CodeT5~\cite{Feng2020} generate vector representations that capture structural and semantic properties of code. These models can be finetuned for downstream tasks such as clone detection. These methods allow models to leverage large amounts of labeled code, typically using contrastive objectives to bring semantically similar fragments closer in embedding space while pushing dissimilar fragments apart~\cite{Bui2021}. However, contrastive approaches require careful selection of negative samples, which can introduce biases and complicate training~\cite{Chen2022}.

Non-contrastive representation learning provides an alternative by enforcing constraints on the embedding space itself, without the need for negative pairs~\cite{Balestriero2022}. Variance-Invariance-Covariance Regularization (VICReg)~\cite{Bardes2022} is a prominent example, optimizing three complementary properties: (i) \emph{invariance} between different views of the same sample, (ii) \emph{variance} across embedding dimensions to prevent collapse, and (iii) \emph{covariance} between features to capture diverse aspects of the input. These properties make VICReg particularly well-suited for Type-IV clone detection, where semantic equivalence exists despite syntactic differences.

Building on this idea, we propose LWVIC4Code, a non-contrastive representation learning approach specifically designed for Type-IV clone detection. Inspired by the layer-wise VICReg formulation of Datta et al.~\cite{Datta2025}, our approach applies VICReg objectives at multiple transformer layers and optimizes them jointly through end-to-end training. Building on this foundation, we introduce a cross-layer consistency regularizer and depth-dependent layer weighting to encourage the progressive refinement of semantic information throughout the network. This leads to the introduction of three novel components as part of the LWVIC4Code architecture. The resulting representations are robust and discriminative, making them suitable for detecting semantically equivalent yet syntactically diverse code fragments. 

We conduct an empirical study to evaluate the effectiveness of LWVIC4Code for Type-IV clone detection, comparing it with a contrastive learning model (CodeBERT$_{CL}$~\cite{kitsios2025detecting}) and zero-shot LLMs. Specifically, we investigate (i) whether LWVIC4Code achieves comparable or superior performance without negative samples, (ii) the benefits of layer-wise supervision over traditional VICReg, (iii) which aspects of the novel components contribute the most to performance gains, and (iv) the generalization of models trained on Python-only data to other programming languages. Our evaluation uses two benchmark datasets: \emph{Kamino}, a large Python-only dataset of Type-IV clones~\cite{Marchezan2026Kamino}, and \emph{GPTCloneBench}, a smaller multi-language dataset covering Python, Java, C\#, and C~\cite{Alam2023}. Performance is assessed using $F_1$-score and Matthews Correlation Coefficient (MCC)~\cite{Chicco2020}.
Statistical significance is evaluated using p-values corrected for multiple comparisons via the Benjamini-Hochberg procedure~\cite{Benjamini1995}.

Results demonstrate that LWVIC4Code achieves strong performance for Type-IV clone detection across multiple programming languages and datasets. For example, on GPTCloneBench, LWVIC4Code reaches $F_1$ scores of 0.977 on C\#, 0.968 on Java, and 0.920 on Python, with corresponding MCC values of 0.957, 0.938, and 0.840. These results outperform the contrastive baseline CodeBERT$_{CL}$ (e.g., $F_1$ 0.899 and MCC 0.794 on C\#) while avoiding the need for negative sampling. Overall, LWVIC4Code achieves higher performance than the original VICReg objective on GPTCloneBench ($p<0.05$), with the ablation study demonstrating that both layer-wise weighting and cross-layer consistency contribute to improving representation quality. Despite this, the last-layer-only variant achieves better results in specific scenarios.

In addition, LWVIC4Code outperforms zero-shot LLM-based approaches, achieving significantly higher performance than GPT-OSS and Qwen and comparable performance to DeepSeek-r1. These results demonstrate that a dedicated non-contrastive representation learning approach can provide more reliable and efficient semantic code representations than general-purpose LLMs for Type-IV clone detection, while requiring substantially less inference cost.
The main contributions of this paper are: 

\begin{leftbar}
       i) We introduce \emph{VIC4Code}, a non-contrastive representation learning approach that adapts the VICReg objective to code representation learning, enabling the learning of semantic embeddings without requiring negative samples;
       
       ii) Building on prior layer-wise VICReg formulations, we propose \emph{LWVIC4Code}, a transformer-based approach for Type-IV clone detection that introduces cross-layer consistency regularization and depth-dependent layer weighting, enabling end-to-end learning of hierarchical code representations and improving the robustness and discriminative power of code embeddings;
   
    iii) A \emph{systematic empirical evaluation of LWVIC4Code for Type-IV clone detection}, demonstrating that it outperforms a state-of-the-art contrastive learning baseline, improves over the standard VICReg objective, and achieves competitive or superior performance compared to zero-shot LLM-based approaches across multiple programming languages and datasets;
    
   iv) An \emph{ablation study of the LWVIC4Code components}, demonstrating that layer-wise weighting and cross-layer consistency are important contributors to performance improvements, with layer-wise weighting providing the largest impact;
   
    v) An \emph{analysis of the impact of training data composition for LWVIC4Code}, comparing Python-only vvs. multi-language training, demonstrating that representations learned from Python transfer effectively to Java and C\#, though less to syntactically divergent languages such as C;
    
    vi) A \emph{set of models for Type-IV clone detection}: two variants of the complete LWVIC4Code (Python-only and multi-language), three variants without each novel component point, and one variant of VIC4Code.
\end{leftbar}

The remainder of the paper is structured as follows. \secref{sec:background} presents background on semantic clone detection and representation learning; \secref{sec:approach} describes the model architectures and non-contrastive training strategies; \secref{sec:evaluation} details the experimental protocol and results; \secref{sec:discussion} provides analysis and limitations; \secref{sec:related} reviews related work; and \secref{sec:conclusion} concludes the paper.
\section{Background}\label{sec:background}
In this section, we outline the main challenges motivating our work, introducing key related concepts.
\subsection{Challenges and Motivation}

Over the years, different definitions and classifications of code clones have been proposed~\cite{Carter1993,Chanchal2009}. The most widely accepted taxonomy distinguishes four types of clones:  
\begin{enumerate*}[label=(\arabic*), itemjoin={{; }}, itemjoin*={{; and }}]
    \item \textit{Types~I--III (textual and syntactic clones):} near-identical or slightly modified code fragments that share substantial syntactic similarity, ranging from exact copies to variations through insertions or deletions
    \item \textit{Type-IV (semantic clones):} functionally equivalent fragments that may differ entirely in syntax or structure, making them the most challenging to detect.
\end{enumerate*}

Type-IV clones are difficult to identify because their similarity lies in program \emph{behavior} rather than structure. \figref{fig:type4-example} illustrates two Python implementations of a function that computes the row sums of a matrix. Although both return the same result given the same input, their syntactic representations share almost no overlap (e.g., one uses nested loops while the other uses `map`), making them hard to detect using traditional text- or token-based techniques~\cite{Wang2023}.
Although this example is relatively simple to detect, real-world clones can be far more complex and require significant time and effort to identify, resources that companies could otherwise invest in more productive activities~\cite{Deissenboeck2010}.


\begin{figure}[h]
  \centering
  \begin{subfigure}[t]{0.46\linewidth}
    \begin{lstlisting}[style=pythonstyle]
def row_sums(matrix):
    sums = []
    for row in matrix:
        total = 0
        for val in row:
            total += val
        sums.append(total)
    return sums
    \end{lstlisting}
    \caption{Primary Solution}
    \label{fig:functionA}
  \end{subfigure}
  \hfill
  \begin{subfigure}[t]{0.52\linewidth}
    \begin{lstlisting}[style=pythonstyle]
def row_sums(matrix):
    return list(
        map(
            sum, matrix
            )
        )

        
    \end{lstlisting}
    \caption{Type-IV clone}
    \label{fig:functionB}
  \end{subfigure}
  \vspace{-1mm}
  \caption{Two functionally equivalent but syntactically different functions in Python}
  \label{fig:type4-example}
\end{figure}

The increasing use of LLMs for code generation~\cite{Murali2024} further exacerbates these challenges, as generated code often introduces semantically equivalent implementations that differ substantially in structure~\cite{Xu2025}. This shift moves clone creation from explicit copy-paste reuse to implicit, model-generated duplication, which can propagate at scale across projects and developers. Consequently, Type-IV clone detection becomes even more critical in modern development practice~\cite{Wang2025}.

Early work explored semantic clone detection through program transformations and behavioral analysis~\cite{Gabel2008}. More recently, ML techniques have been adopted to capture deeper semantic relationships between code fragments~\cite{Bhaskar2024,Sheneamer2016}. These approaches typically rely on learning vector representations of code, known as \emph{code embeddings}, which aim to capture structural and semantic properties of programs~\cite{chen2019literature}. Once learned, such embeddings can be used to measure similarity between code fragments and identify potential clones~\cite{buch2019learning, defreez2018path}. 

Despite these advances, accurately detecting Type-IV clones using traditional methods remains challenging~\cite{Zhang2023}, as semantic similarity can manifest through diverse implementation strategies, control-flow structures, or library usage patterns~\cite{Gabel2008}. Rule-based or token-level approaches often fail to capture these high-level behavioral patterns. Consequently, representation learning is a promising direction, as it enables models to automatically transform raw code into structured, informative, and low-dimensional feature spaces~\cite{Bengio2013}. This allows models to capture semantic properties of programs beyond superficial syntactic similarity. In this context, representation learning has emerged as an effective paradigm for learning such representations without requiring large amounts of labeled data, as demonstrated in domains such as computer vision~\cite{Datta2025}.

In source code representation learning, however, contrastive learning methods have become popular~\cite{Bui2021,kitsios2025detecting,li2023zc}. These methods bring semantically related code fragments closer in the embedding space while pushing unrelated fragments apart~\cite{Bui2021}. However, contrastive approaches require large numbers of negative samples and careful sampling strategies, which can introduce biases and increase training complexity~\cite{Chen2022}. Moreover, selecting negative examples for semantic clone detection is non-trivial. Common strategies include using Types~I--III clones as negatives~\cite{Alam2023}, choosing syntactically similar but semantically different fragments~\cite{Li2026}, or selecting completely unrelated code~\cite{Svajlenko2015}. Each approach has trade-offs, as overly dissimilar negatives may lead models to rely on trivial cues, while syntactically similar negatives require careful curation and may lead to overfitting.

An alternative approach is \emph{non-contrastive learning}~\cite{Balestriero2022}. Rather than relying on negatives, these methods enforce constraints on the embedding space itself to learn invariant representations. Variance-Invariance-Covariance Regularization (VICReg)~\cite{Bardes2022}, originally proposed for vision tasks, encourages representations to remain invariant across different views while maintaining variance and decorrelation. These properties are particularly appealing for Type-IV clones, where behavioral equivalence may manifest through diverse syntax, making non-contrastive learning a promising candidate for semantic code analysis~\cite{zakeri2023systematic}.

Despite their benefits, non-contrastive methods require a large amount of quality data for effective training and validation. Traditional benchmarks such as \textit{BigCloneBench}~\cite{Svajlenko2022} are primarily syntactic, lack balanced negatives, and do not fully capture Type-IV behavior~\cite{Krinke2022,Krinke2025}. Other large-scale datasets, such as \emph{CodeNet}~\cite{CodeNet}, \emph{CodeSearchNet}~\cite{Husain2020}, and \emph{BigCodeBench}~\cite{Zhuo2024bigcodebench}, have significantly advanced research in ML for code. These datasets support tasks such as code generation, summarization, and retrieval. However, they were not specifically designed for clone detection, and therefore lack explicit labels or guarantees of semantic equivalence between code fragments.

More recently, datasets targeting Type-IV clones have begun to emerge. For example, \textit{\emph{GPTCloneBench}}~\cite{Alam2023} contains Type-IV clones across multiple programming languages, including Python, Java, C\#, and C. The dataset was generated by prompting GPT with code fragments from \emph{SemanticCloneBench}~\cite{al2020semanticclonebench} and subsequently refined through manual curation, tool-assisted filtering, functionality testing, and automated validation. While valuable, \emph{GPTCloneBench} remains relatively small for Type-IV clones ($\approx 25k$ pairs across all programming languages) and may not provide sufficient diversity to train robust representation learning models. Another prominent dataset, \emph{Kamino}~\cite{Marchezan2026Kamino}, was derived using multiple LLMs to generate code from BigCodeBench entries, followed by extensive testing and filtering. Kamino contains a larger number of Type-IV clones ($\approx 79k$ pairs), making it particularly suitable for training non-contrastive representation learning methods.

Finally, clone detection is a practical, development-oriented challenge~\cite{Kim2018}. Solutions must be suitable for real-time detection during development. Thus, clone detection mechanisms based on LLMs are typically resource-intensive, whereas lightweight embedding-based models, such as representation learning, offer faster, more efficient detection and easier integration into IDEs. Furthermore, while LLMs provide promising capabilities for semantic reasoning over code, their effectiveness for clone detection remains an active area of research~\cite{kitsios2025detecting,Moumoula2025}.


In summary, two main challenges motivate this study:
\begin{enumerate*}[label=(\arabic*), itemjoin={{; }}, itemjoin*={{; and }}]
    \item \textit{Learning semantic representations of code:} detecting Type-IV clones requires models that capture behavioral equivalence across highly diverse implementations, a challenge amplified by the current growing use of LLM-generated code in practice
    \item \textit{Limitations of contrastive learning:} contrastive methods depend on high-quality negative samples, which are difficult to define for semantic clones and can introduce bias and additional training complexity.
\end{enumerate*}

These challenges motivate the need for a non-contrastive-based approach that can learn robust semantic representations without relying on negative sampling.

\subsection{Code Embeddings for Clone Detection}\label{sec:finetuned} 

Pretrained models such as \emph{CodeBERT}~\cite{Wang2021} and \emph{CodeT5}~\cite{Feng2020} generate code embeddings by training on large code corpora using self-supervised objectives like masked language modeling. These embeddings capture both structural and semantic patterns in code, making them a strong foundation for various code understanding tasks. In the context of Type-IV clone detection, embeddings are typically used as feature representations for code fragments. Given a pair of code snippets $c_1$ and $c_2$, the model produces embeddings $z_1$ and $z_2$. The similarity between these embeddings is commonly computed using cosine similarity:

\begin{equation}
\text{sim}(z_1, z_2) = \frac{z_1 \cdot z_2}{\|z_1\|\|z_2\|}.
\end{equation}

A threshold $\theta$ is then applied to determine whether the pair constitutes a clone:

\begin{equation}
(c_1, c_2) \text{ is a clone if } \text{sim}(z_1, z_2) \ge \theta.
\end{equation}

Choosing an appropriate $\theta$ is critical, as it directly affects false positives and false negatives. In practice, it is often selected empirically by evaluating performance metrics (e.g., $F_1$, MCC) across a range of thresholds or using percentile-based approaches on similarity distributions~\cite{Li2026}. 

Pretrained embeddings can be finetuned on labeled clone datasets to improve performance, either using single-language Type-IV clones~\cite{Marchezan2026Kamino} or combining multiple languages~\cite{Alam2023,kitsios2025detecting}. However, such approaches typically rely on contrastive signals or explicitly labeled negative examples to distinguish semantically similar from dissimilar code. For Type-IV clones, constructing high-quality negatives is particularly challenging and may introduce bias or degrade performance. These limitations motivate the exploration of non-contrastive methods, which learn robust semantic representations without requiring negative examples, as discussed in the next section.
\section{Non-Contrastive Representation Learning for Type-IV Clone Detection}\label{sec:approach}

This section describes the non-contrastive representation learning strategies proposed in this work for detecting Type-IV code clones. 
Our approach builds on VICReg~\cite{Bardes2022} to capture code semantics through embeddings, adapting it from its original focus on image embeddings.

\subsection{VIC4Code: Non-Contrastive Representation Learning}

VIC4Code is a transformer-based encoder that leverages non-contrastive representation learning to capture semantic similarity in Type-IV clones. The model operates on two views (i.e., two clones) of the same code fragment, enforcing three complementary constraints on the embedding space: \begin{enumerate*}[label=(\roman*), itemjoin={{; }}, itemjoin*={{; and }}]
    \item \textbf{Invariance:} intuitively, embeddings of semantically equivalent code fragments should be close, even if their syntax differs
    \item \textbf{Variance:} each embedding dimension should maintain variability to distinguish non-clones, preventing collapse
    \item \textbf{Covariance:} embedding dimensions should capture complementary semantic features, reducing redundancy
\end{enumerate*}
Formally, let $Z = \{x_1, x_2\}$ denote the projected embeddings of a batch of code fragment pairs. The VICReg loss combines three complementary terms:
\begin{equation}
\mathcal{L} =
\lambda \mathcal{L}_{\text{inv}} +
\mu \mathcal{L}_{\text{var}} +
\nu \mathcal{L}_{\text{cov}},
\end{equation}
where each term corresponds to one of the constraints (invariance, variance, and covariance) and is weighted by hyperparameters $(\lambda, \mu, \nu)$. The encoder $E$ produces latent embeddings $z$, which are projected via a learnable projector $P$ into a space suitable for VICReg optimization. In addition, the final-layer representation is obtained using the CLS token embedding produced by the encoder.

The training procedure of VIC4Code is summarized in \algoref{alg:VIC4Code}. The encoder $E$ (line 1) is a transformer-based model with a tokenizer capable of converting raw source code into embeddings, such as CodeBERT~\cite{Wang2021}. Along with the learnable projector $P$ and the optimizer, the model is initialized before training begins. For each epoch (lines 2--10), batches of code fragment pairs $(c_1, c_2)$ are sampled from the dataset $\mathcal{D}$ (line 3). Each fragment is encoded into latent embeddings $z_1$ and $z_2$ by the encoder (line 4), and subsequently projected into a higher-dimensional space via $P$ to obtain $x_1$ and $x_2$ (line 5). The VICReg loss $\mathcal{L}$ is then computed over the projected embeddings (line 6), enforcing the invariance, variance, and covariance constraints. This loss is backpropagated through both the encoder and projector to update all model parameters (line 7). After all epochs are completed, the final encoder weights are saved (line 11) and can be used to generate embeddings for downstream Type-IV clone detection tasks.

\begin{algorithm}[t]
\caption{VIC4Code for Type-IV clone detection}
\label{alg:VIC4Code}
\begin{algorithmic}[1]
\State Initialize encoder $E$, projector $P$, and optimizer
\For{epoch = $1$ to $N_{\text{epochs}}$}
    \For{each batch $(c_1, c_2) \sim \mathcal{D}$}
        \State Encode: $z_1 = E(c_1),\; z_2 = E(c_2)$
        \State Project: $x_1 = P(z_1),\; x_2 = P(z_2)$
        \State Compute VICReg loss $\mathcal{L}$
        \State Backpropagate $\mathcal{L}$ and update parameters
    \EndFor
\EndFor
\State Save final encoder weights
\end{algorithmic}
\end{algorithm}

\subsection{LWVIC4Code: Layer-Wise VICReg}
Applying the VICReg loss only at the final encoder layer, as in VIC4Code, may miss semantic information captured at intermediate layers. Transformer-based encoders encode hierarchical representations~\cite{Datta2025}. In this context, early layers capture syntactic patterns, while deeper layers represent higher-level semantics. This observation motivates our layer-wise VICReg approach, which propagates the loss across multiple depths to encourage semantically meaningful embeddings at every layer.

Inspired by the layer-wise VICReg formulation proposed by Datta et al.~\cite{Datta2025}, we extend VIC4Code by applying VICReg objectives at multiple transformer depths. However, unlike the original layer-wise formulation, which optimizes layers using local objectives and layer-local gradient updates, LWVIC4Code is trained end-to-end, allowing gradients from all layer-wise objectives to jointly update the encoder parameters, while the stop-gradient operation is restricted to the previous-layer representation within the cross-layer regularizer. Building on this foundation, we introduce two additional mechanisms specifically designed for code representation learning: (i) a cross-layer consistency regularizer that encourages smooth semantic refinement across layers and (ii) depth-dependent weighting that progressively emphasizes higher-level semantic representations captured by deeper layers.

\begin{figure}[t]
    \centering
    \includegraphics[width=\columnwidth]{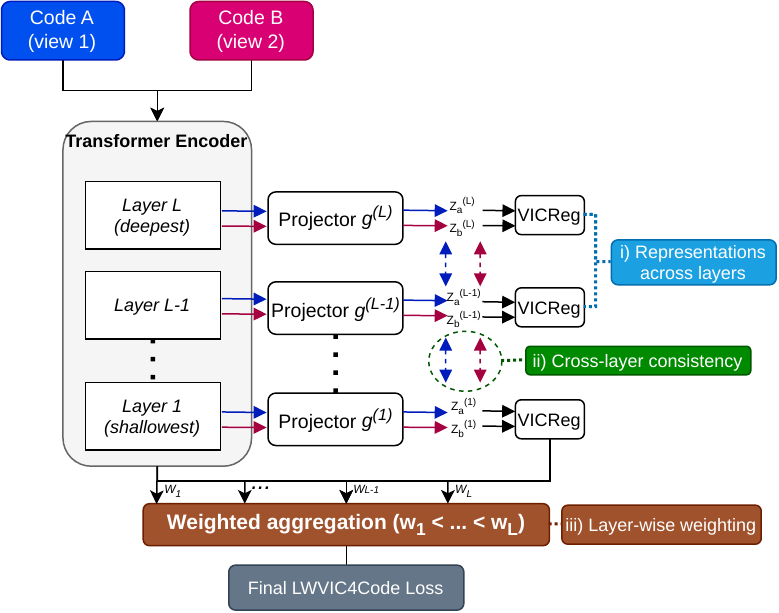}
    \caption{LWVIC4Code architecture.}
    \label{fig:architecture}
\end{figure}

In summary, we extend VIC4Code by: (i) optimizing representations across all transformer layers; (ii) enforcing cross-layer consistency; and (iii) applying layer-wise weighting. \figref{fig:architecture} illustrates the overall LWVIC4Code architecture, including the three novel architectural components that lead to the final LWVIC4Code loss calculation. Each component point is described next.

\subsubsection{Optimizing representations across layers}

Let $h^{(l)} \in \mathbb{R}^{T \times d}$ denote the hidden representation produced by transformer layer $l$, where $T$ is the number of input tokens and $d$ is the hidden dimension. Unlike VIC4Code, which relies on the final-layer CLS representation, LWVIC4Code applies masked mean pooling at every transformer layer to obtain layer-specific representations:

\begin{equation}
p^{(l)} =
\frac{\sum_{t=1}^{T} m_t h_t^{(l)}}{\sum_{t=1}^{T} m_t},
\end{equation}

where $m_t$ denotes the attention mask indicating valid tokens. This operation excludes padding tokens while preserving information from all code tokens. A layer-specific projector $g^{(l)}$ then maps the pooled representation to the embedding space $z^{(l)}$:

\begin{equation}
z^{(l)} = g^{(l)}(p^{(l)}).
\end{equation}

At each layer, the VICReg loss encourages three properties in the embedding space: invariance between semantically similar code fragments, sufficient variance across dimensions to prevent collapse, and reduced covariance between dimensions to capture complementary features:

\begin{equation}
\mathcal{L}^{(l)} =
\lambda \mathcal{L}_{\text{inv}}^{(l)} +
\mu \mathcal{L}_{\text{var}}^{(l)} +
\nu \mathcal{L}_{\text{cov}}^{(l)}.
\end{equation}

\subsubsection{Enforcing cross-layer consistency}

To ensure that semantic information accumulated in earlier layers is not discarded during optimization, we introduce a cross-layer consistency term. The objective is not to make consecutive layer representations identical, but rather to encourage semantic continuity across the hierarchy. In other words, deeper layers should refine and enrich the information learned by previous layers while preserving the core semantic content of the code fragment. Without such regularization, layer-specific VICReg objectives may drive adjacent layers toward substantially different embedding spaces, reducing the benefits of hierarchical representation learning. Therefore, if a code fragment's embedding at layer $l-1$ encodes certain semantic features, the embedding at layer $l$ should refine these features rather than producing a completely unrelated vector. 

The cross-layer consistency term is computed independently for both views in a clone pair  and then averaged:
\begin{equation}
\mathcal{L}_{\text{cross}}^{(l)} = \frac{1}{2}\left(
\left\| z_a^{(l)} - \text{sg}(z_a^{(l-1)}) \right\|_2^2 +
\left\| z_b^{(l)} - \text{sg}(z_b^{(l-1)}) \right\|_2^2
\right),
\end{equation}
where $\text{sg}(\cdot)$ denotes the stop-gradient operator, and $z_a^{(l)}, z_b^{(l)}$ are the embeddings of the two code fragments at layer $l$ (see ``2.~Cross-layer consistency'' in \figref{fig:architecture})). We employ an L2 consistency objective because it directly penalizes large representation shifts between adjacent layers while remaining computationally inexpensive and stable to optimize. Unlike cosine-based objectives, which preserve only directional similarity, the L2 formulation constrains both direction and magnitude of the embedding trajectory. 

The cross-layer consistency loss acts as an auxiliary regularizer at each transformer depth. To jointly optimize semantic alignment within each layer and semantic continuity across layers, we combine the layer-wise VICReg objective and the cross-layer consistency term into a single training objective:

\begin{equation}
\mathcal{L}_{\text{LayerVICReg}} =
\sum_{l=1}^{L} w_l \left( \mathcal{L}^{(l)} + \alpha \mathcal{L}_{\text{cross}}^{(l)} \right),
\end{equation}

where $w_l = (l/L)^2$ emphasizes deeper layers and $\alpha$ controls the cross-layer consistency strength.

\begin{algorithm}[t]
\caption{LWVIC4Code for Type-IV clone detection}
\label{alg:layer_VIC4Code}
\begin{algorithmic}[1]
\State Initialize encoder $E$, layer-wise projectors $g^{(l)}$, and optimizer
\For{epoch = $1$ to $N_{\text{epochs}}$}
    \For{each batch $(c_1, c_2) \sim \mathcal{D}$}
        \State Initialize hidden states $h_1^{(0)}, h_2^{(0)}$
        \For{layer $l = 1$ to $L$}
            \State Forward pass: $h_1^{(l)}, h_2^{(l)}$
            \State Masked mean pooling:
$p_1^{(l)},p_2^{(l)}=\mathrm{MeanPool}(h_1^{(l)},h_2^{(l)},m_1,m_2)$
            \State Project: $z_1^{(l)} = g^{(l)}(p_1^{(l)}),\; z_2^{(l)} = g^{(l)}(p_2^{(l)})$
            \State Compute $\mathcal{L}^{(l)}$
            \If{$l > 1$}
                \State Compute cross-layer term: $\mathcal{L}_{\text{cross}}^{(l)} = \|z^{(l)} - \text{sg}(z^{(l-1)})\|_2^2$
            \EndIf
            \State Accumulate weighted loss $w_l (\mathcal{L}^{(l)} + \alpha \mathcal{L}_{\text{cross}}^{(l)})$
        \EndFor
        \State Backpropagate total loss and update parameters
    \EndFor
\EndFor
\State Save final encoder weights
\end{algorithmic}
\end{algorithm}

LWVIC4Code training procedure is summarized in Algorithm~\ref{alg:layer_VIC4Code}. 
Similar to VIC4Code, it requires an encoder capable of transforming raw code fragments into embedding representations, such as a pretrained tokenizer model (e.g., CodeBERT) (line~1). 
For each training epoch (line~2), batches of code fragment pairs $(c_1, c_2)$ are processed (line~3). 
The hidden states for both code fragments are initialized at the input layer (line~4), and then a forward pass is performed through each layer $l$ of the encoder (line~5). 
At every layer, token-level representations are aggregated using masked mean pooling, where padding tokens are ignored according to the attention mask, producing fixed-size vectors $p_1^{(l)}$ and $p_2^{(l)}$ (line~7). 

\subsubsection{Applying layer-wise weighting}
 The VICReg loss is computed independently at each layer (line~9). For layers beyond the first, an additional cross-layer consistency term $\mathcal{L}_{\text{cross}}^{(l)}$ is calculated to encourage smooth transitions between consecutive layers, where the stop-gradient operation prevents the cross-layer term from directly updating the previous-layer representation, avoiding trivial solutions, while the remaining objectives continue to optimize the encoder end-to-end. (lines~10--11).  
 The combined layer-wise and cross-layer loss is scaled by a depth-dependent weight $w_l$ and accumulated into the total loss (line~12). Once all layers are processed, the total loss is backpropagated to update both the encoder and all layer-specific projectors (line~13). After completing all epochs, the final encoder weights are saved (line~14) for downstream Type-IV clone detection tasks (see ``3.~Layer-wise weighting'' in \figref{fig:architecture}).

Although the cross-layer consistency objective introduces an additional attraction force between consecutive layer representations, the variance and covariance components inherited from VICReg provide opposing regularization pressures that discourage collapse. The variance term encourages sufficient activation variability across embedding dimensions, whereas the covariance term promotes complementary feature representations. Consequently, the cross-layer objective encourages semantic continuity across the transformer hierarchy while the VICReg regularization terms maintain representation diversity. Moreover, the VICReg-based approach is less sensitive to batch composition than contrastive methods (\secref{sec:finetuned}), since variance and covariance regularization prevent representation collapse without requiring negatives.

\section{Empirical Evaluation}\label{sec:evaluation}

The goal of our evaluation is to empirically assess the effectiveness and generalizability of non-contrastive representation learning (LWVIC4Code) for detecting Type-IV code clones. Specifically, we aim to determine whether LWVIC4Code can (i) achieve performance comparable to contrastive approaches and zero-shot LLMs, (ii) benefit from a layer-wise training strategy to improve semantic representation learning, and (iii) generalize across programming languages when trained on a single programming language.

\subsection{Evaluation Protocol}

To guide our empirical study, we define the following research questions (RQ):

\textbf{RQ1:} To what extent does non-contrastive representation learning (LWVIC4Code) compare to state-of-the-art contrastive learning for Type-IV clone detection?

\emph{Rationale}: We aim to assess whether LWVIC4Code can achieve performance comparable to a state-of-the-art contrastive learning approach while avoiding the need for negative samples.

\emph{Method}: We compare LWVIC4Code against CodeBERT$_{CL}$~\cite{kitsios2025detecting}, a recent state-of-the-art contrastive learning model for semantic clone detection. CodeBERT$_{CL}$ replaces the original classifier of CodeBERT with a contrastive learning objective that minimizes the distance between clone pairs while enforcing a margin between non-clone pairs. Similarity is computed using cosine similarity over the learned representations. During training, CodeBERT$_{CL}$ requires both positive (clone) and negative (non-clone) pairs, where negative pairs are randomly sampled to maintain a balanced 1:1 ratio. In contrast, LWVIC4Code is trained exclusively using positive clone pairs through the VICReg objective, eliminating the need for negative sampling. Both approaches are trained and evaluated on the Kamino (Python-only, 80/20 train-test split) and GPTCloneBench (Python, Java, C\#, and C, 40/60 train-test split) datasets.

\emph{Metrics}: Performance is evaluated using precision ($P$), recall ($R$), F1-score ($F_1$), and MCC, as it is a more reliable metric for binary classification because it considers all parts of the confusion matrix and avoids inflated performance~\cite{Chicco2020}. For each pair of code embeddings, predictions are obtained by applying a similarity threshold $\tau$. Thresholds are explored in the range $[0.1, 1.0]$ with increments of $0.05$, and for each setting, the threshold maximizing MCC is selected (breaking ties using $F_1$). This approach ensures that each model is compared at its optimal operating point, providing a fair assessment of the results. 
To assess statistical significance, we model the relationship between similarity scores and ground truth labels using Generalized Estimating Equations (GEE) logistic regression~\cite{Hardin2002}, which accounts for repeated measurements at the case level. 
Discriminative performance is quantified via Area Under the Curve (AUC)~\cite{Bradley1997}, with differences estimated using clustered bootstrap~\cite{Christman2000} to account for data dependencies. All p-values from bootstrap comparisons are adjusted using the Benjamini-Hochberg procedure~\cite{Benjamini1995} to control the false discovery rate. Differences are considered statistically significant if the adjusted p-value is below 0.05. Analyses are reported per dataset, per language, and overall to ensure robust conclusions.

\vspace{0.5em}
\textbf{RQ2:} To what extent does a layer-wise training strategy (LWVIC4Code) improve the effectiveness of VIC4Code for Type-IV clone detection?

\emph{Rationale:} VIC4Code optimizes embeddings only at the final layer, which may miss intermediate semantic features. LWVIC4Code enforces VICReg constraints across all layers, capturing richer semantic representations. We aim to measure whether this layer-wise learning improves alignment between semantically similar code fragments for Type-IV clones.
\emph{Method}: VIC4Code and LWVIC4Code are compared under identical training conditions. Both models are trained on the Kamino and GPTCloneBench datasets using the same encoder architecture, optimization settings, and VICReg coefficients $(\lambda=34, \ \mu=36, \ \nu=1)$. These coefficient values were selected through iterative testing on dataset samples to maximize $F_1$ and MCC performance.
\emph{Metrics}: We use the same evaluation metrics ($P$, $R$, $F_1$, MCC) and statistical significance tests as in RQ1.

\vspace{0.5em}
\textbf{RQ3:} To what extent each layer-wise component of LWVIC4Code contributes to Type-IV clone detection?

\emph{Rationale}: LWVIC4Code extends VIC4Code through three design choices: (i) optimizing representations across all transformer layers. (ii) enforcing cross-layer consistency, and (iii) applying layer-wise weighting. We aim to quantify the contribution of each component and identify which components are primarily responsible for the observed performance improvements.

\emph{Method}: We perform an ablation study by comparing the complete LWVIC4Code model against three variants, each removing a single component while keeping all other training settings unchanged:
(i) \textit{Last Layer Only} (LWVIC4Code$_{LL}$), which computes the VICReg objective exclusively on the final transformer layer;
(ii) \textit{No Cross-Layer Consistency} (LWVIC4Code$_{NC}$), which disables the consistency objective between adjacent transformer layers; and (iii) \textit{No Layer-wise Weighting} (LWVIC4Code$_{NW}$), which assigns equal importance to every transformer layer during optimization.
All variants are trained using the same datasets, encoder architectures, optimization settings, and coefficients.

\emph{Metrics}: We evaluate the same metrics used in the previous RQs (precision, recall, $F_1$, MCC, and AUC), together with the same statistical significance analysis for each variant.

\vspace{0.5em}
\textbf{RQ4:} To what extent does LWVIC4Code compare to zero-shot LLMs for Type-IV clone detection?

\emph{Rationale:} We aim to determine whether LWVIC4Code provides competitive or superior performance to LLMs in detecting semantic clones, without the overhead of zero-shot prompting.
\emph{Method:} We evaluate LWVIC4Code against zero-shot LLMs (\texttt{deepseek-r1:14b}, \texttt{gpt-oss:20b}, \texttt{qwen3.5:9.65b}). A fixed prompt (see \tabref{tab:prompt}) is used to extract predicted labels.  No finetuning is applied. While larger LLMs could be explored, we focus on models that are publicly accessible and representative of current practice.
\emph{Metrics}: Evaluation is performed using the same metrics as in the previous RQs (precision, recall, $F_1$, MCC). These metrics, however, are calculated based on the predicted labels (binary task) rather than using cosine similarity like RQs 1-2.  For this reason, we use McNemar's test to assess statistically significant differences between paired classification outcomes using the best threshold for LWVIC4Code to determine a prediction and compare it with the LLM's predicted result. Similar to the other RQs, all p-values are adjusted using the Benjamini-Hochberg procedure~\cite{Benjamini1995} to control the false discovery rate, and differences are considered statistically significant if the adjusted p-value is below 0.05. 

\begin{table}[t]
\centering
\footnotesize
\caption{LLM Prompt for Type-IV Clone Detection.}
\vspace{-1mm}
\begin{tabular}{p{0.95\linewidth}} 
\toprule
\textbf{System Prompt:} \\
You are a careful code reviewer. Determine whether two functions are Type-IV (semantic) clones. Output only a JSON object:
\{\texttt{"prediction": 0 or 1}\}. No explanations. \\\midrule
\textbf{User Prompt:} \\
Given two code snippets (\texttt{code\_a}, \texttt{code\_b}) in a target language, determine if they are semantic clones. 
Prediction = 1 if the functions are semantic clones; otherwise, 0\\
\bottomrule
\end{tabular}\vspace{-1mm}
\label{tab:prompt}
\end{table}

\vspace{0.5em}
\textbf{RQ5:} To what extent can representations learned from Python-only generalize to other programming languages?

\emph{Rationale:} We investigate the transferability of semantic embeddings to assess whether LWVIC4Code trained on a single language can detect Type-IV clones in other languages.
\emph{Method:} LWVIC4Code is trained exclusively on Python Type-IV clones from Kamino. Evaluation is performed on Java, C\#, and C clones from GPTCloneBench. We examine performance across all possible cosine similarity thresholds $\tau \in [0.0, 1.0]$ to assess sensitivity to threshold selection.
\emph{Metrics}: We consider MCC and $F_1$ scores for all possible cosine similarity thresholds, ranging $[0.0, 1.0]$ to observe how the different thresholds affect results.

\subsection{Datasets}

\paragraph{Kamino Dataset.}
The Kamino dataset~\cite{Marchezan2026Kamino} is a large-scale collection of semantically equivalent Python code fragments derived from BigCodeBench~\cite{Zhuo2024bigcodebench}, comprising $78,771$ Type-IV clone pairs.
The clones were generated via a hybrid pipeline that produces behaviorally equivalent but syntactically diverse implementations with LLMs, followed by deterministic validation steps. The pipeline ensures semantic correctness through unit test execution, enforces syntactic diversity using CodeBLEU-based filtering, and selects non-redundant examples via clustering. We use an 80/20 train-test split, following standard practice. Training data is used for RQs 1--3, while the test split is used in all evaluations. 

\paragraph{GPTCloneBench.} 
GPTCloneBench~\cite{Alam2023} is a multi-language dataset containing Types I--IV clones. The dataset was created by leveraging code fragments from SemanticCloneBench~\cite{al2020semanticclonebench} and prompting GPT-3 to generate semantically equivalent implementations, followed by filtering out syntactic clones using NiCad and manual validation. For this study, we use only the Type-IV subset, which includes code fragments across Python ($3\,948$ pairs), Java ($9\,935$ pairs), C\# ($6\,872$ pairs), and C ($4\,823$ pairs). Due to its smaller size, we use a 40/60 train-test split to ensure sufficient test coverage while retaining minimal training data for multi-language evaluation. Training data is used in RQs 1--2, while the test split is used in all RQs.

\subsection{Results}\label{sec:results}

In this section, we present the empirical results for each RQ. The approach and complete results are available in our replication package~\cite{ASE2026ReplicationPackage}.

\subsubsection{RQ1: LWVIC4Code vs Contrastive Learning}
LWVIC4Code consistently outperforms the state-of-the-art contrastive learning approach, CodeBERT$_{CL}$~\cite{kitsios2025detecting}, while requiring only positive clone pairs during training. Across all datasets and programming languages (\tabref{tab:results}), LWVIC4Code achieves higher precision, recall, $F_1$, and MCC scores. For example, on GPTCloneBench (\texttt{C\#}), LWVIC4Code achieves an $F_1$ of 0.977 and an MCC of 0.957, substantially outperforming CodeBERT$_{CL}$ ($F_1=0.900$, MCC = 0.794). Similar improvements are observed for \texttt{Java} ($F_1$: 0.968 vs. 0.804; MCC: 0.938 vs. 0.612) and \texttt{Python} ($F_1$: 0.920 vs. 0.777; MCC: 0.840 vs. 0.598). The largest performance gap occurs for \texttt{C}, where LWVIC4Code achieves an $F_1$ of 0.866 and an MCC of 0.790, compared to only $F_1=0.570$ and MCC = 0.457 for CodeBERT$_{CL}$. On the Python-only Kamino dataset, LWVIC4Code also provides a clear improvement, increasing the $F_1$ score from 0.873 to 0.926 and the MCC from 0.745 to 0.853.

Looking at overall discriminative ability using AUC, LWVIC4Code consistently outperforms CodeBERT$_{CL}$ across all datasets and programming languages. On GPTCloneBench, LWVIC4Code achieves an overall AUC of 0.9864, compared with 0.8881 for CodeBERT$_{CL}$, corresponding to an improvement of 0.0983 (95\% CI [0.0938, 0.1031], $p<0.001$). The largest gain is observed for \texttt{C}, where LWVIC4Code improves the AUC from 0.7812 to 0.9612. 
Substantial improvements are also obtained for \texttt{Java} (0.9891 vs.\ 0.8905),
\texttt{Python} (0.9704 vs.\ 0.8864)
, and \texttt{C\#} (0.9953 vs.\ 0.9597)
, all with corrected $p<0.001$. On the Python-only Kamino dataset, LWVIC4Code similarly achieves a higher AUC (0.9751) than CodeBERT$_{CL}$ (0.9326), corresponding to an improvement of 0.0425 (95\% CI [0.0402, 0.0448], $p<0.001$). All reported p-values were adjusted using the Benjamini--Hochberg procedure and remain below 0.05.

\begin{leftbar}
\textbf{RQ1:} LWVIC4Code consistently outperforms CodeBERT$_{CL}$, across all evaluated datasets and programming languages, improving the overall AUC from 0.8881 to 0.9864 on GPTCloneBench and from 0.9326 to 0.9751 on Kamino, with corresponding gains of up to 0.297 in $F_1$ and 0.333 in MCC ($p<0.001$). 
\end{leftbar}

\subsubsection{RQ2: Layer-wise vs Traditional VIC4Code}

Across datasets and languages, LWVIC4Code consistently improves over standard VIC4Code (\tabref{tab:results}). On GPTCloneBench, LWVIC4Code achieves its highest $F_1$ and MCC on \texttt{C\#} (0.977 and 0.957), surpassing VIC4Code (0.965 $F_1$, 0.933 MCC). Improvements are also observed for \texttt{Java} (MCC 0.938 vs. 0.921) and \texttt{Python} (MCC 0.840 vs. 0.837). Even for the lowest-performing language (\texttt{C}), LWVIC4Code outperforms VIC4Code (MCC 0.790 vs. 0.741). All p-values remain below 0.05 after being corrected using the Benjamini-Hochberg procedure.

Considering overall discriminative ability, LWVIC4Code achieves slightly higher AUC values on GPTCloneBench (0.9864 vs. 0.9815) with the largest gains for \texttt{C}, while differences are smaller for other languages. On Python-only Kamino, VIC4Code performs marginally better (AUC 0.9775 vs. 0.9751), but both methods remain highly effective.  

\begin{leftbar}
\textbf{RQ2:} LWVIC4Code consistently improves over VIC4Code for Type-IV clone detection across programming languages, achieving higher $F_1$ and MCC scores on GPTCloneBench, with gains of up to 0.049 MCC on \texttt{C} and 0.024 MCC on \texttt{C\#} ($p<0.05$). LWVIC4Code also achieves a higher overall AUC on GPTCloneBench (0.9864 vs. 0.9815), while both approaches remain highly effective on Kamino. 
\end{leftbar}

\begin{table}[t]
\centering
\footnotesize
\addtolength{\tabcolsep}{-2pt}
\caption{Clone detection results for best thresholds ($\theta$). Bold indicates the best result within each dataset/language group, while underlined indicates the second-best result.}
\vspace{-1mm}
\begin{tabular}{L{2cm}L{.5cm}L{.7cm}R{.6cm}|R{.6cm}R{.6cm}R{.6cm}R{.8cm}}
\toprule
\textbf{Model} & \textbf{DS} & \textbf{Lang.} & $\boldsymbol{\theta}$ & \textbf{$P$} & \textbf{$R$} & \textbf{$F_1$} & \textbf{MCC} \\
\midrule
\multicolumn{8}{c}{\textbf{Contrastive Baseline}}\\
\midrule
\multirow{4}{*}{CodeBERT$_{CL}$} & \multirow{4}{*}{G} & Java & .75 &	.813 &	.795 &	.804&	.612 \\
 &  & C\# & .55 &	.866&	.935&	.899&	.794 \\
 &  & C & .90 &	.929 &	.411 &	.570 &	.457 \\
 &  & Py & .85	& .853 &	.714 &	.777 &	.598 \\ \hline
 & \multirow{1}{*}{K} & Py & .70&	.872&	.874&	.873&	.745 \\
\midrule
\multicolumn{8}{c}{\textbf{Non-contrastive methods}}\\\midrule
\multirow{4}{*}{LWVIC4Code} & \multirow{4}{*}{G} & Java & .75 & .952 & .985 & \underline{.968} & \underline{.938} \\
 &  & C\# & .75 & .967 & .988 & \underline{.977} & \underline{.957} \\
 &  & C & .90 & .879 & .854 & .866 & .790 \\
 &  & Py & .80 & .919 & .921 & .920 & .840 \\ \hline
 & \multirow{1}{*}{K} & Py & .70 & .941 & .910 & .926 & .853 \\
\midrule
\multirow{4}{*}{VIC4Code} & \multirow{4}{*}{G} & Java & .90 & .949 & .970 & .960 & .921 \\
 &  & C\# & .85 & .954 & .977 & .965 & .933 \\
 &  & C & .90 & .787 & .900 & .840 & .741 \\
 &  & Py & .75 & .912 & .926 & .919 & .837 \\ \hline
 & \multirow{1}{*}{K} & Py & .60 & .959 & .934 & .946 & .894 \\
\midrule
\multicolumn{8}{c}{\textbf{LWVIC4Code Ablation}}\\
\midrule
\multirow{4}{*}{LWVIC4Code$_{LL}$} & \multirow{4}{*}{G} & Java & .80 & .967 & .979 & \textbf{.973} & \textbf{.945} \\
 &  & C\# & .90 & .989 & .975 & \textbf{.982} & \textbf{.964} \\
 &  & C & .90 & .939 & .917 & \textbf{.928} & \textbf{.857} \\
 &  & Py & .80 & .964 & .943 & \textbf{.953} & \textbf{.908} \\ \hline
 & \multirow{1}{*}{K} & Py & .75 & .979 & .933 & \underline{.956} & \underline{.915} \\
\midrule
\multirow{4}{*}{LWVIC4Code$_{NC}$} & \multirow{4}{*}{G} & Java & .80 & .956 & .976 & .966 & .931 \\
 &  & C\# & .85 & .976 & .974 & .975 & .950 \\
 &  & C & .80 & .894 & .944 & \underline{.918} & \underline{.834} \\
 &  & Py & .85 & .934 & .929 & \underline{.932} & \underline{.863} \\ \hline
 & \multirow{1}{*}{K} & Py & .80 & .955 & .889 & .921 & .849 \\
\midrule
\multirow{4}{*}{LWVIC4Code$_{NW}$} & \multirow{4}{*}{G} & Java & .80 & .936 & .942 & .939 & .877 \\
 &  & C\# & .75 & .968 & .968 & .968 & .936 \\
 &  & C & .85 & .868 & .937 & .901 & .796 \\
 &  & Py & .70 & .816 & .937 & .872 & .734 \\ \hline
 & \multirow{1}{*}{K} & Py & .70 & .941 & .849 & .893 & .799 \\
\midrule
\multicolumn{8}{c}{\textbf{LLMs zero-shot}}\\\midrule
\multirow{4}{*}{Deepseek-r1} & \multirow{4}{*}{G} & Java & -- & .992 & .789 & .879 & .805 \\
 &  & C\# & -- & .998 & .820 & .900 & .837 \\
 &  & C & -- & .993 & .558 & .715 & .662 \\
 &  & Py & -- & .994 & .639 & .778 & .681 \\ \hline
 & \multirow{1}{*}{K} & Py & -- & 1 & .916 & \textbf{.956} & \textbf{.919} \\
\midrule
\multirow{4}{*}{Gpt-oss} & \multirow{4}{*}{G} & Java & -- & .991 & .695 & .817 & .727 \\
 &  & C\# & -- & .999 & .724 & .840 & .759 \\
 &  & C & -- & .991 & .412 & .582 & .549 \\
 &  & Py & -- & .984 & .537 & .694 & .593 \\ \hline
 & \multirow{1}{*}{K} & Py & -- & .999 & .776 & .873 & .795 \\
\midrule
\multirow{4}{*}{Qwen} & \multirow{4}{*}{G} & Java & -- & .994 & .531 & .692 & .603 \\
 &  & C\# & -- & .999 & .585 & .738 & .651 \\
 &  & C & -- & .996 & .302 & .464 & .461 \\
 &  & Py & -- & 1 & .390 & .561 & .493 \\ \hline
 & \multirow{1}{*}{K} & Py & -- & 1 & .306 & .469 & .425 \\
\bottomrule
\multicolumn{8}{c}{G: GPTCloneBench; K: Kamino}
\end{tabular}\vspace{-1mm}
\label{tab:results}
\end{table}
\subsubsection{RQ3: Ablation on LWVIC4Code}

The ablation results demonstrate the importance of the proposed three novel components in LWVIC4Code (\tabref{tab:results}). Surprisingly, the last-layer-only variant achieves the highest $F_1$ and MCC scores among the evaluated variants on GPTCloneBench. This suggests that, for this dataset, optimizing only the final transformer layer is sufficient to achieve strong discriminative performance, although the reasons for this behavior require further investigation. For example, on \texttt{C\#}, LWVIC4Code$_{LL}$ improves MCC from 0.957 to 0.964, while on \texttt{Java} it increases MCC from 0.938 to 0.945.

The remaining ablation variants show that the additional layer-wise mechanisms contribute positively to the overall representation quality. Removing cross-layer consistency (LWVIC4Code$_{NC}$) results in small improvements in isolated cases, such as \texttt{C} on GPTCloneBench ($F_1$ 0.918 vs. 0.866 for the original LWVIC4Code) and \texttt{C\#} ($F_1$ 0.975 vs. 0.977), but generally decreases performance compared to the complete model. Similarly, removing layer-wise weighting (LWVIC4Code$_{NW}$) leads to performance degradation in most settings, with notable reductions for \texttt{Python} on GPTCloneBench (MCC 0.734) and \texttt{Kamino} (MCC 0.799). Although some configurations remain competitive, these results indicate that assigning equal importance to all layers is less effective than the proposed weighting strategy, suggesting that different transformer layers contribute unequally to the final representation.

The results show that removing layer-wise weighting (LWVIC4Code$_{NW}$) consistently degrades performance, with statistically significant AUC reductions on both GPTCloneBench (0.9864 vs. 0.9701, $p<0.001$) and Kamino (0.9751 vs. 0.9481, $p<0.001$). Similarly, removing cross-layer consistency produces significant differences on Kamino ($p<0.001$), although the AUC difference is smaller (0.9751 vs. 0.9684). For the last-layer-only variant, the results vary across datasets, as it achieves higher AUC on GPTCloneBench (0.9916 vs. 0.9864, $p=0.001$), whereas LWVIC4Code achieves higher AUC on Kamino (0.9840 vs. 0.9751, $p=0.001$). This difference is also reflected in the threshold-based evaluation, where LWVIC4Code$_{LL}$ performs better than LWVIC4Code for higher $\theta$ values, as indicated by the best thresholds reported in \tabref{tab:results}, while it is less effective for lower thresholds ($\theta$<0.75). 

\begin{figure}[t]
    \centering
    \includegraphics[width=\linewidth]{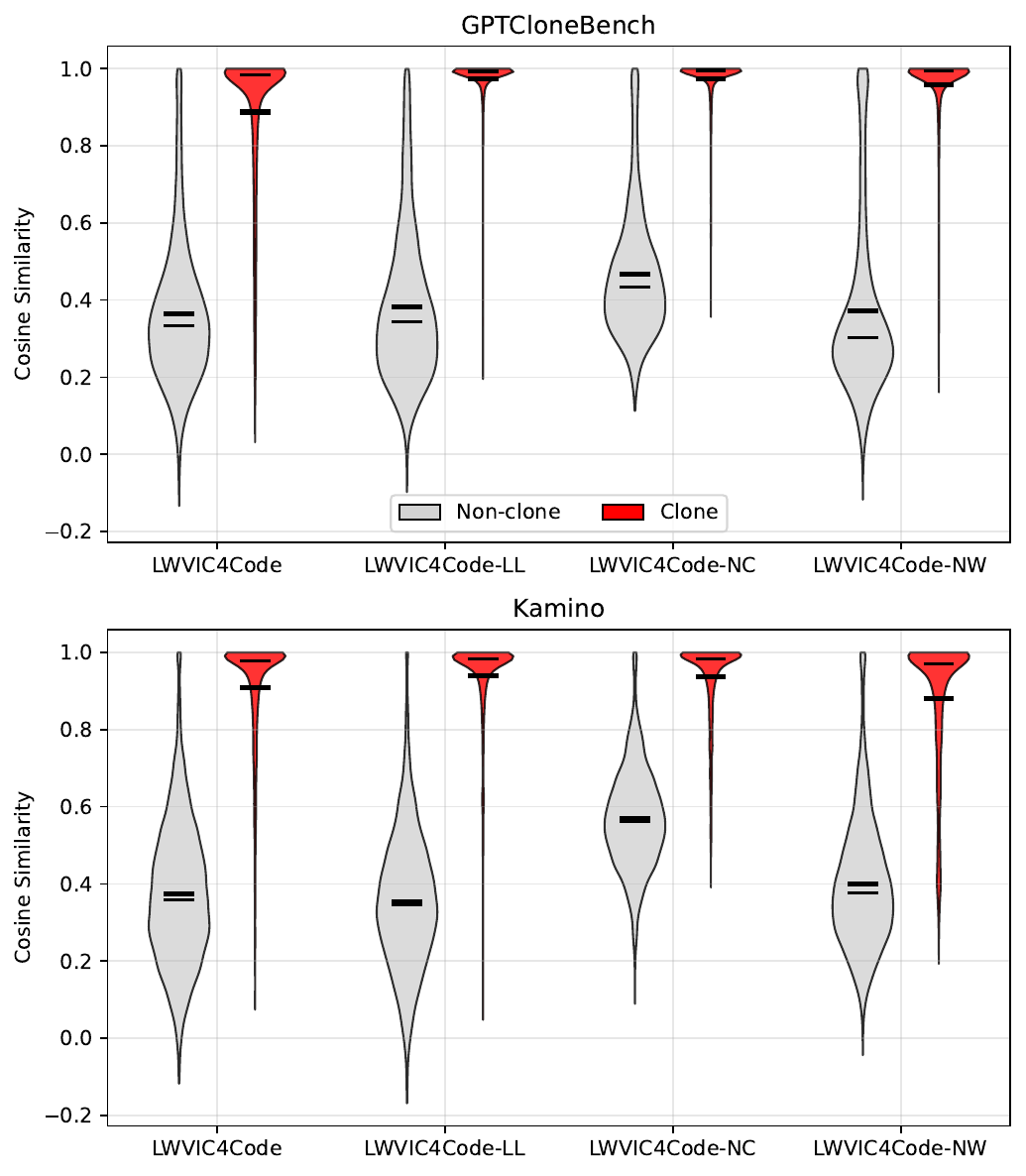}
    \caption{Cosine similarity distributions of clone and non-clone pairs for LWVIC4Code variants.}
    \label{fig:rq3violin}
\end{figure}

Lastly, an analysis of the similarity score distributions (see \figref{fig:rq3violin}) further shows that LWVIC4Code$_{LL}$ produces consistently higher median similarity scores for clone pairs than the complete model across both GPTCloneBench (0.993 vs. 0.983) and Kamino (0.983 vs. 0.977), while also exhibiting lower variance, particularly on GPTCloneBench (std. 0.063 vs. 0.190). These findings explain why LWVIC4Code$_{LL}$ reaches its optimal performance at higher similarity thresholds, although the full LWVIC4Code model remains more robust across datasets and operating thresholds.

\begin{leftbar}
\textbf{RQ3:} LWVIC4Code benefits from its novel components, with layer-wise weighting providing the most consistent improvement for Type-IV clone detection. Removing cross-layer consistency also affects performance, although with smaller and dataset-dependent effects. The last-layer-only variant achieves competitive results and even higher AUC on GPTCloneBench (0.9916 vs. 0.9864), but performs worse on Kamino (0.9751 vs. 0.9840).
\end{leftbar}

\subsubsection{RQ4: LWVIC4Code vs Zero-Shot LLMs}

Across programming languages, LWVIC4Code generally outperforms zero-shot LLMs for Type-IV clone detection (\tabref{tab:results}). On GPTCloneBench, LWVIC4Code achieves more consistent results across languages, particularly for structurally diverse ones like \texttt{C}, where LLM performance drops sharply (e.g., DeepSeek-r1 MCC is 0.662). The only exception where the LLM outperforms LWVIC4Code is DeepSeek-r1 for Kamino (0.956 $F_1$, 0.919 MCC). These results may be good considering that Kamino used Deepseek-r1 as one of the LLMs to generate its clones. Most of the other cases (GPTCloneBench) showed average to poor performance, with the best results being for DeepSeek-r1 on \texttt{C\#} (0.900 $F_1$, 0.837 MCC) and on \texttt{Java} (0.879 $F_1$, 0.805 MCC)

Hence, for all LLMs evaluated on GPTCloneBench, McNemar’s tests indicate statistically significant differences compared to LWVIC4Code after Benjamini–Hochberg correction ($p_{adj}<0.05$). The largest differences are observed for Qwen and GPT-OSS. On the Kamino dataset, the statistical differences are also significant for all evaluated LLMs. However, the performance gap is smaller for DeepSeek-r1, whereas GPT-OSS and Qwen remain below. 

\begin{leftbar}
\textbf{RQ4:} On GPTCloneBench, LWVIC4Code achieves up to 0.316 MCC improvement over zero-shot LLMs (e.g., compared to Qwen on \texttt{C}) ($p<0.05$). On Kamino, DeepSeek-r1 is the only LLM that surpasses LWVIC4Code ($F_1=0.956$, MCC$=0.919$). 
\end{leftbar}

\subsubsection{RQ5: Transferability of Representations in Python}

\begin{figure}
    \centering
    \includegraphics[width=0.9\linewidth]{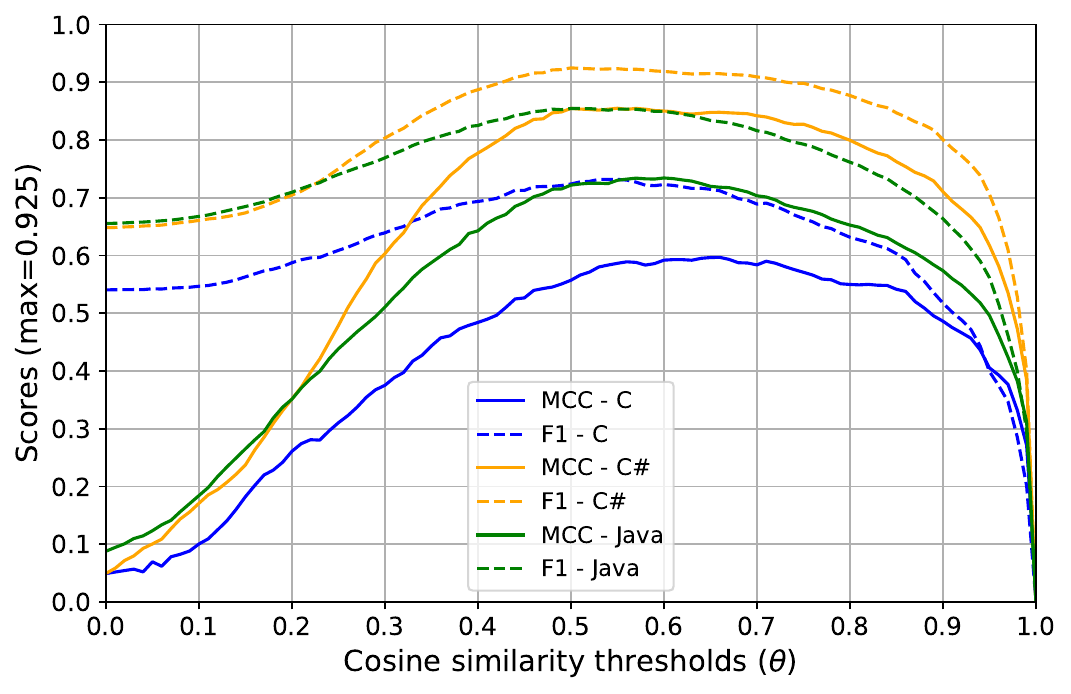}\vspace{-2mm}
    \caption{Scores across  $\theta$ using Python-trained LWVIC4Code.}\vspace{-3mm}
    \label{fig:rq5}
\end{figure}

\figref{fig:rq5} presents the MCC and $F_1$ scores across thresholds when LWVIC4Code is trained exclusively on Python Type-IV clones from Kamino. Overall, the results demonstrate a strong degree of multi-language generalization, particularly for \texttt{C\#} and \texttt{Java}.

For these two languages, the model achieves its best performance in the threshold range $\theta \in [0.6, 0.7]$, where $F_1$ scores exceed 0.8 and peak at 0.925 for \texttt{C\#}. Performance remains relatively stable up to $\theta = 0.75$, after which both MCC and $F_1$ decline, indicating that higher thresholds impose stricter similarity requirements that the model cannot satisfy.

In contrast, performance on \texttt{C} is noticeably lower, with a maximum $F_1$ slightly above 0.7 and MCC reaching approximately 0.6. This gap can be attributed to the greater syntactic and paradigmatic differences between \texttt{C} and Python, compared to the more structurally similar \texttt{C\#} and \texttt{Java}.
\begin{leftbar}
\textbf{RQ5:} LWVIC4Code, trained solely on Python data, can effectively transfer to other programming languages without any exposure during training. This indicates that the model captures language-agnostic semantic representations of code, although its effectiveness decreases as the target language diverges further from Python (i.e., on \texttt{C}).
\end{leftbar} 

\section{Discussion}\label{sec:discussion}
\subsection{Benefits and Limitations}
\paragraph{Advantages of LWVIC4Code}
LWVIC4Code exhibits better performance than the contrastive baseline. LWVIC4Code achieved high $F_1$ and MCC scores on GPTCloneBench for Java and C\#, as shown in \tabref{tab:results} (0.968 and 0.938 for Java, and 0.977 and 0.957 for C\#) even though the training sets for these languages were relatively small ($4\,257$ and $2\,944$ pairs, respectively). Its non-contrastive nature further highlights its efficiency, as negative pairs were not needed to achieve competitive performance, and in some cases, it outperforms zero-shot LLMs such as DeepSeek-r1, GPT-OSS, and Qwen, particularly on structurally diverse languages like C ($F_1$ 0.866, MCC 0.790) and Python ($F_1$ 0.920, MCC 0.840). These results demonstrate that LWVIC4Code can learn robust, language-agnostic semantic representations of code efficiently, even from limited data. These findings directly address the challenges outlined in~\secref{sec:background}, demonstrating that effective semantic representations can be learned without reliance on negative sampling.

\paragraph{Contribution of LWVIC4Code novel components}
The ablation study (RQ3) provides further insights into the design choices behind LWVIC4Code. Removing layer-wise weighting consistently reduces performance across datasets, indicating that not all encoder layers contribute equally to semantic clone representations. These findings are consistent with prior work~\cite{Datta2025} suggesting that deeper transformer layers encode more abstract semantic information. Similarly, removing cross-layer consistency leads to significant performance degradation on Kamino, demonstrating that explicitly encouraging agreement between representations learned at different depths improves the stability of the embedding space.
Interestingly, the last-layer-only variant achieves competitive, and in some cases superior, AUC values, particularly on GPTCloneBench. However, this improvement is dependent on the similarity threshold ($\theta$), as the variant performs better only for higher thresholds, while LWVIC4Code provides stronger results across a broader range of thresholds. This indicates that relying exclusively on the final encoder layer may produce highly concentrated similarity scores that are effective in stricter decision scenarios. In contrast, the complete LWVIC4Code architecture provides more stable representations, which may generalize to more scenarios where lower thresholds are intended (e.g., cross-language clone detection).

\paragraph{Limitations of non-contrastive learning}
Since the model relies solely on positive pairs, it requires a sufficiently large and diverse set of examples to avoid collapsing representations toward overly similar embeddings. This is reflected in the evaluation results: on languages with fewer or more syntactically divergent examples (e.g., C), LWVIC4Code achieves lower $F_1$ and MCC scores (0.866 and 0.790, respectively) and requires higher thresholds ($\theta = 0.9$) to maintain precision, indicating a bias toward high similarity values. In contrast, contrastive models naturally balance positives and negatives, which helps maintain a wider separation between clone and non-clone embeddings. Moreover, while LWVIC4Code transfers well to languages structurally similar to Python (Java, C\#), its effectiveness diminishes as the target language diverges, suggesting that non-contrastive representations may be less robust when the training data does not capture sufficient cross-language diversity. 


\paragraph{Practical implications}
LWVIC4Code is a lightweight and efficient model, making it well-suited for real-time integration into development environments such as IDEs. It is also efficient, as detecting clones across the full GPTCloneBench test set ($\approx 44k$ pairs) takes less than 10 minutes, whereas zero-shot LLMs require dozens of hours to process the same data. Moreover, LWVIC4Code can be finetuned incrementally on new positive examples in production without generating negative samples. In contrast, contrastive approaches require careful construction of negative pairs, which can be time-consuming and impractical in real-world settings. These properties make LWVIC4Code a practical solution for continuous, scalable, and language-agnostic Type-IV clone detection in real-world development environments.

\paragraph{Use of LLM-generated data}
A key aspect of our study is the use of LLM-generated data, as Kamino consists of Python code produced by LLMs. While this may lead to threats, such a choice reflects a growing reality, as much modern code is now produced or assisted by LLMs in practice~\cite{Tabarsi2025}. Evaluating LWVIC4Code and other models on such data provides insights into their effectiveness in realistic AI-assisted coding scenarios. Importantly, GPTCloneBench (created by leveraging SemanticCloneBench fragments with GPT-3, followed by manual curation, tool-assisted filtering, functionality testing, and automated validation) contains high-quality human-labeled and cross-language code. Experiments on this dataset show that findings from LLM-generated data transfer effectively to realistic code, supporting the practical relevance of our results in modern AI-assisted development settings.

\paragraph{Research opportunities}
Our findings highlight several opportunities for future work. First, exploring hybrid training strategies that combine non-contrastive and contrastive objectives could help mitigate limitations related to a lack of data and representation collapse, particularly for syntactically divergent languages (e.g., Python vs C). Second, extending LWVIC4Code to additional programming languages and more diverse datasets (including code generated by different LLMs or real-world industrial repositories) would further assess its generalizability. Third, integrating dynamic or adaptive threshold selection mechanisms could enhance practical deployment, reducing reliance on manually tuned $\theta$ values (e.g., allowing one to use LWVIC4Code$_{LL}$ instead). Finally, investigating lightweight incremental finetuning and continual learning strategies at production time would allow the model to adapt efficiently to evolving codebases and AI-assisted development environments, supporting scalable, real-time clone detection. More broadly, as this work represents, to the best of our knowledge, the first application of VICReg to code-related tasks, it opens a new line of research on non-contrastive representation learning for software engineering. Beyond clone detection, such methods could be explored for related problems, including code search, defect detection, and program classification, where learning robust semantic representations without reliance on negative sampling may offer significant advantages.

\subsection{Threats to Validity}

\emph{Internal validity:} Differences in model performance for contrastive, non-contrastive, and LLMs may be affected by implementation details, hyperparameter choices, or dataset preprocessing. We mitigated this by using consistent training settings, architectures, and evaluation protocols across all models and approaches. Additionally, we selected a state-of-the-art baseline~\cite{kitsios2025detecting} that used a well-established pretrained model (CodeBERT) for contrastive learning and carefully tuned hyperparameters for LWVIC4Code based on preliminary experiments to ensure fair and representative comparisons. To ensure comparability with the contrastive model, we used CodeBERT as the encoder for both LWVIC4Code and VIC4Code, maintaining consistency in the embedding space across all approaches.

\emph{External validity:} Our results may not fully generalize to other codebases or programming languages. To mitigate this threat, we evaluated models on two complementary datasets: Kamino, a large-scale Python-only dataset, and GPTCloneBench, a multi-language dataset covering four programming languages. Although Kamino is derived from BigCodeBench and may include multiple implementations per task, cross-split semantic similarity on task descriptions is low (mean = 0.16, std = 0.14), indicating mostly distinct functionalities. Lastly, Kamino is composed of LLM-generated Python code, which could introduce biases or patterns not present in real-world code. To mitigate this, we also evaluate on GPTCloneBench, a high-quality, curated dataset with cross-language and human-labeled clones. Consistent results across both datasets suggest that our findings generalize beyond synthetic LLM-generated data.

\emph{Construct validity:} Metrics such as MCC, $F_1$, and similarity thresholds may not capture all aspects of semantic equivalence. Furthermore, selecting one (random) $\theta$ could bias the results analysis towards an improvement in a selected direction. To mitigate this, we considered the results for all possible $\theta$, reporting the best for each individual model and programming language. Additionally, prior work has shown that clone detection models such as CodeBERT can achieve artificially high performance when test data is similar to training data~\cite{kitsios2025detecting, sonnekalb2022generalizability}. By including multiple languages and ensuring non-overlapping train-test splits, we reduce this risk and provide a more realistic assessment of model generalization. Furthermore, the results for RQ5 were considered by training and testing on completely different datasets, demonstrating the potential of transfer learning.
 
\emph{Conclusion validity:} While MCC and $F_1$ are well-known metrics, only considering them could bias the analysis. To mitigate this, we also report the statistical significance tests (GEE, AUC differences with bootstrap confidence intervals, McNemar's test) to reduce the risk of drawing incorrect inferences. We also applied the Benjamini–Hochberg~\cite{Benjamini1995} procedure to control the false discovery rate across multiple comparisons. However, small sample sizes in some language subsets (e.g., C) may affect the reliability of these results.

\section{Related Work}\label{sec:related}

This section presents an overview of the main approaches proposed for semantic/Type-IV code clone detection. 
\subsection{AST and PDG-Based Approaches}


Sheneamer and Kalita~\cite{Sheneamer2016} combined Abstract Syntax Tree (AST) and Program Dependence Graphs (PDG)-based representations with ML classifiers to detect complex Type-III and Type-IV clones. Their results showed that semantic program representations significantly outperform traditional metric-based approaches, highlighting the importance of incorporating program semantics into clone detection. Recent work~\cite{zou2020ccgraph} leveraged PDG representations with graph analysis techniques to detect semantic clones, reporting strong performance on benchmark datasets. Despite their effectiveness, AST- and PDG-based approaches often suffer from scalability limitations, as constructing and analyzing detailed program graphs can be computationally expensive, making them difficult to apply to large codebases~\cite{wu2022detecting}.

Furthermore, since our approach is inspired by the layer-wise VICReg~\cite{Datta2025}, it is important to highlight what is novel in LWVIC4Code. LWVIC4Code differs from the layer-wise VICReg formulation of Datta et al. in both its application domain (code clone detection vs. image classification) and optimization strategy. Datta et al.'s approach uses layer-wise representation learning on local objectives applied independently at each layer. In contrast, LWVIC4Code applies layer-wise VICReg objectives within a transformer encoder for source code and optimizes the entire network jointly through standard end-to-end backpropagation (``1. Representation across layers'' in \figref{fig:architecture}). The proposed cross-layer consistency regularizer (``2. Cross-layer consistency'' in \figref{fig:architecture}) and depth-dependent weighting (``3. Layer-wise weighting'') further encourage semantic alignment across layers and prioritize deeper semantic representations relevant to Type-IV clone detection.

\subsection{Representation Learning Approaches}

To address scalability limitations, recent work has explored representation learning techniques. Wu et al.~\cite{wu2022detecting} proposed a method that models semantic code clones by transforming AST structures into Markov chain representations to train a clone detection model, while Hu et al. ~\cite{hu2022treecen} instead convert AST structures into graph-based representations to learn semantic embeddings of code. These approaches demonstrate how structural program information can be transformed into representations suitable for machine learning models.

Considering contrastive representation learning, Kitsios et al.~\cite{kitsios2025detecting} demonstrate that contrastive objectives help models generalize to functionalities not observed during training.
Similarly, Li et al.~\cite{li2023zc} proposed a cross-language code clone detection approach that learns embeddings from multiple clone datasets. Their model is able to generalize to unseen programming languages and features by learning language-agnostic semantic representations. This is similar to what was explored with RQ5, which showcases how our LWVIC4Code can be trained in Python only to detect clones in Java and C\#. 
Mohammed et al.~\cite{mohammed2025cross} proposed a transformer-based model for cross-language semantic clone detection that incorporates semantic hints derived from AST and control-flow graph structures, achieving strong results across multiple languages.

In a different direction, Li et al.~\cite{li2025bega} apply contrastive learning to binary code embeddings rather than source code. Although primarily designed for security applications such as binary similarity detection, their results further highlight the effectiveness of contrastive representation learning for capturing program semantics.
A limitation of contrastive approaches, however, lies in their dependence on carefully constructed positive and negative pairs (as discussed in~\secref{sec:background}). Dataset imbalance can significantly impact model performance. For example, modifying the BigCloneBench dataset to balance positive and negative pairs may introduce potential biases during training~\cite{kitsios2025detecting}.

To the best of our knowledge, no prior work has employed a non-contrastive objective to train models for code clone detection; alternative training strategies beyond purely contrastive approaches have been explored~\cite{Zhang2025}. For instance, Keller et al.~\cite{keller2021you} proposed a visualization-based method that transforms source code into image-like representations and leverages transfer learning from computer vision models. By exploiting pretrained visual feature extractors, their approach captures semantic patterns in code; however, it still relies on labeled data. Similarly, Guo et al.~\cite{Guo2020} introduced a semi-supervised clone detection technique that facilitates review sharing. Their method integrates a convolutional neural network with an autoencoder architecture to learn from both labeled and unlabeled data, thereby improving detection performance.

More recently, Dou et al.~\cite{Dou2026} propose an ensemble learning approach for automated code clone verification, combining multiple machine learning models to improve the accuracy of clone classification. Their work, however, focuses on verifying candidate clone pairs rather than learning semantic code representations, relying on engineered features and ensemble strategies to distinguish true clones from false positives.

\subsection{Large Language Model-Based Approaches}

LLMs have been explored for semantic clone detection due to their strong capability to capture contextual information in code. Zhang et al.~\cite{zhang2024assessing} evaluated the ability of GPT-3.5 and GPT-4 to detect semantic clones and found that these models struggled with accurately identifying functional equivalence between programs. Recently, however, several off-the-shelf open-source LLMs perform on par with specialized clone detection models when evaluating functionalities not present during training, highlighting the potential of these models for generalization~\cite{kitsios2025detecting}. Thus, justifying the comparison to LWVIC4Code in RQ4. 
Nevertheless, challenges remain. In the context of cross-language clone detection, Moumoula et al.~\cite{Moumoula2025} show that even advanced LLMs struggle to reliably capture functional equivalence across programming languages. Their findings suggest that dedicated embedding models yield more robust semantic representations and currently achieve state-of-the-art performance for cross-lingual clone detection. This observation aligns with our findings for RQ4, where LLMs achieve competitive results only at lower $\theta$ thresholds, yet still underperform compared to both contrastive and non-contrastive embedding models reported in~\tabref{tab:results}.


\section{Conclusion}\label{sec:conclusion}

This paper presents LWVIC4Code, a non-contrastive representation learning approach for Type-IV code clone detection that adapts the VICReg framework to code and enforces layer-wise consistency across transformer layers. By capturing semantic information progressively, LWVIC4Code produces robust embeddings capable of detecting semantically equivalent, syntactically diverse code fragments without relying on negative samples. 
Our evaluation demonstrates LWVIC4Code's strong performance across multiple programming languages, achieving better $F_1$ and MCC scores relative to a contrastive baseline and zero-shot LLMs. The ablation study showed that removing layer-wise weighting has the biggest negative impact on the model. Despite this, using the last layer only showed improvements for higher $\theta$. 
Our evaluation also shows that when trained on Python-only data, it can generalize effectively to Java and C\#, highlighting its potential for cross-language transfer.
Overall, this work contributes a new class of models for Type-IV clone detection, provides insights into layer-wise non-contrastive learning, and systematically compares different machine learning strategies, laying the foundation for further research on semantic code representation and multi-language clone detection.

Future work includes exploring hybrid training strategies that combine non-contrastive and contrastive objectives, extending LWVIC4Code to additional languages and diverse datasets (including LLM-generated and industrial code), integrating dynamic threshold selection, and investigating incremental fine-tuning and continual learning for scalable, real-time clone detection.

\section*{Acknowledgments}
This work was partially funded by the Natural Sciences and Engineering Research Council of Canada (NSERC), grant numbers RGPIN-2025-05677 and RGPIN-2026-06532.

\section*{Data Availability Statement}
A replication package is available at~\cite{ASE2026ReplicationPackage}.

\bibliographystyle{IEEEtran}
\bibliography{bib/references}

\end{document}